\documentclass{article}

\usepackage{amssymb,amsfonts,amsmath,color}
\usepackage{url}
\usepackage{dblfloatfix} 
\usepackage{graphicx}

\definecolor{lightblue}{RGB}{135,206,250}
\definecolor{darkblue}{RGB}{70,130,180}

\usepackage{geometry}
\makeatletter %
   \renewcommand{\@biblabel}[1]{#1.}
\makeatother

\newcommand{\footremember}[2]{%
    \footnote{#2}
    \newcounter{#1}
    \setcounter{#1}{\value{footnote}}%
}
\newcommand{\footrecall}[1]{%
    \footnotemark[\value{#1}]%
}

\begin{document}

\title{Universality in human cortical folding across lobes of individual brains}

\author{%
Yujiang Wang\footremember{ca}{yujiang.wang@ncl.ac.uk}\footremember{icos}{Interdisciplinary Computing and Complex BioSystems (ICOS), School of Computing, Newcastle University, Newcastle upon Tyne NE4 5TG, UK}\footremember{ion}{Institute of Neuroscience, Newcastle University, Newcastle upon Tyne NE1 7RU, UK}\footremember{ionucl}{Institute of Neurology, University College London, London WC1N 3BG, UK}%
\and Joe Necus\footrecall{icos} \footrecall{ion}%
\and Luis Peraza Rodriguez\footrecall{ion} 
\and Peter Neal Taylor\footrecall{icos} \footrecall{ion} \footrecall{ionucl}%
\and Bruno Mota\footremember{ufrj}{Instituto de F\'{i}sica, Universidade Federal do Rio de Janeiro, Rio de Janeiro, Brazil}}

\date{}
\maketitle

\begin{abstract}
Background: We have previously demonstrated that cortical folding across mammalian species follows a universal scaling law that can be derived from a simple theoretical model. The same scaling law has also been shown to hold across brains of our own species, irrespective of age or sex. These results, however, only relate measures of complete cortical hemispheres. There are known systematic variations in morphology between different brain regions, and region-specific changes with age. It is therefore of interest to extend our analyses to different cortical regions, and analyze the scaling law within an individual brain.

Methods: To directly compare the morphology of sub-divisions of the cortical surface in a size-independent manner, we base our method on a topological invariant of closed surfaces. We reconstruct variables of a complete hemisphere from each lobe of the brain so that it has the same gyrification index, average thickness and average Gaussian curvature.

Results: We show that different lobes are morphologically diverse but obey the same scaling law that was observed across human subjects and across mammalian species. This is also the case for subjects with Alzheimer's disease. The age-dependent offset changes at similar rates for all lobes in healthy subjects, but differs most dramatically in the temporal lobe in Alzheimer's disease.

Significance: Our results further support the idea that while morphological parameters can vary locally across the cortical surface/across subjects of the same species/across species, the processes that drive cortical gyrification are universal.

\end{abstract}

\newpage
\section{Introduction}

Cortical folding, or gyrification, is one of the most striking features that can appear in mammalian brains, with many recent studies investigating its mechanisms \cite{Zilles2013,Tallinen2014,RonanFletcher2014,Bayly2014,Striedter2015,Tallinen2016,Garcia2018}. Based on a simple theoretical model, we recently proposed a universal law that relates average cortical thickness $T$, exposed area $A_e$ and total area $A_t$ of a cerebral cortex, and describes its degree of folding:

\begin{equation}\label{UnivLaw}
  A_t \sqrt{T}= k A_{e}^{\alpha},
\end{equation}
where our model predicts that $\alpha = 5/4$. This model is based on the assumption that cortical morphology minimizes an effective free energy that takes into account a pressure-like term (hypothesized to be exerted by white matter axonal tension and hydrostatic cerebral spinal fluid pressure) and the self avoiding nature of the cortical surface \cite{Mota2015}.

The conceptual significance of this scaling law is that variations in e.g. the exposed area, must be compensated by countervailing changes in thickness or total area. We have observed and reported this in action, first across cortices of different species \cite{Mota2015}, then across different individuals (humans) \cite{Wang2016}. In spite of a great morphological variety in both studies, the quantities co-vary such that the same universal gyrification relation, with the same values for $\alpha=1.25$, is approximately followed in all cases.

The only free parameter is thus $k$, or offset, a dimensionless coefficient that is related to the pressure term in the model. Examining the human data in greater detail, we have shown that there is no significant variation in the scaling exponent $\alpha$ but we noted a systematic decrease of the offset parameter $k$ with age, which may be interpreted as a slackening of white matter axonal tension \cite{Wang2016}.

We can think of both the exponent $\alpha$ and the offset $k$ as two natural probes into cortical morphology, applicable across any number of conditions and groups. The former tests the universality of the gyrification mechanism, and the latter measures changes in physical properties. However, as currently formulated, these probes are only applicable to full cortical hemispheres. It is known that there are systematic variations in both measures across morphological directions, between lobes, and region-specific changes with age (e.g. \cite{Zilles1988,Sowell2003,Sowell2004,Hogstrom2012,Fjell2015}). Furthermore, there are a number of neurological conditions that have been observed to produce morphological anomalies primarily or solely in specific cortical regions (e.g. \cite{Dickerson2009b,Janssen2014,Nesvag2014,Bos2015}). Hence, a natural question is: Are the mechanisms of gyrification the same for all cortical regions? Are there systematic regional differences in the timing and extent of folding during aging, and in health vs. disease? The main purpose of this paper is to generalize our morphological probes to cortical regions so as to be able to answer these questions.

\subsection{Methodological background}

To answer the questions above, two main methodological steps must be taken: The first is to find a consistent way of partitioning cortices. One such natural partition divides each hemisphere into four lobes (frontal, occipital, parietal, and temporal), where the definition of each is tolerably consistent across both individuals and species \cite{Fischl2002,Kaas2013}. The second step is to assess if each partition separately follows the same universal scaling law (Eq.~1) as the whole hemisphere.

One might think that for the second step, simply performing the same scaling analysis on each lobe as was done for the whole cortex would be sufficient. However, a simple thought experiment shows the problem: Let us assume a uniform cortex with constant cortical thickness $T$, is partitioned into segments having the same gyrification index ($g=A_t/A_e$). Then the exposed $A_e$ and total $A_t$ areas of each partition would correspond to the same fraction of their respective values for the full cortex. Thus, in the $x=log(A_e)$ and $y=log(A_t\sqrt{T})$ projection, the data points corresponding to each partition would all line up along a line of slope 1 (signifying constant gyrification index), ordered according to size. A heterogeneous cortex would then introduce further dispersion perpendicular to the line of slope 1 (see a more detailed description in Suppl. Text E). In either case this approach would not capture the scaling law, but only differences in partition size, thickness and gyrification index.

Instead we propose to rely on the known invariance of the integrated Gaussian curvature on a closed surface.\footnote{This result is a consequence of the famous Gauss-Bonnet theorem \cite{Hazewinkel2013encyclopaedia}, which states that the total integrated Gaussian curvature in a closed surface (such as that of cortical hemisphere) is a topological invariant.}
For each partition, we obtain equivalent $A_e'$ and $A_t'$ values that a putative full cortical hemisphere would have, if it had on average the same gyrification index, cortical thickness and average Gaussian curvature for its total and exposed surfaces as the partition. These estimates ($A_e'$ and $A_t'$), together with a small estimation error correction (see Methods for details) can then be used to evaluate if the scaling law (Eq.~\ref{UnivLaw}) also applies across partitions of the same cortex.

\section{Results}

\subsection{Lobes of the same cortex also obey the universal scaling law}

\begin{figure}
\begin{center}
\centerline{\includegraphics[width=18cm]{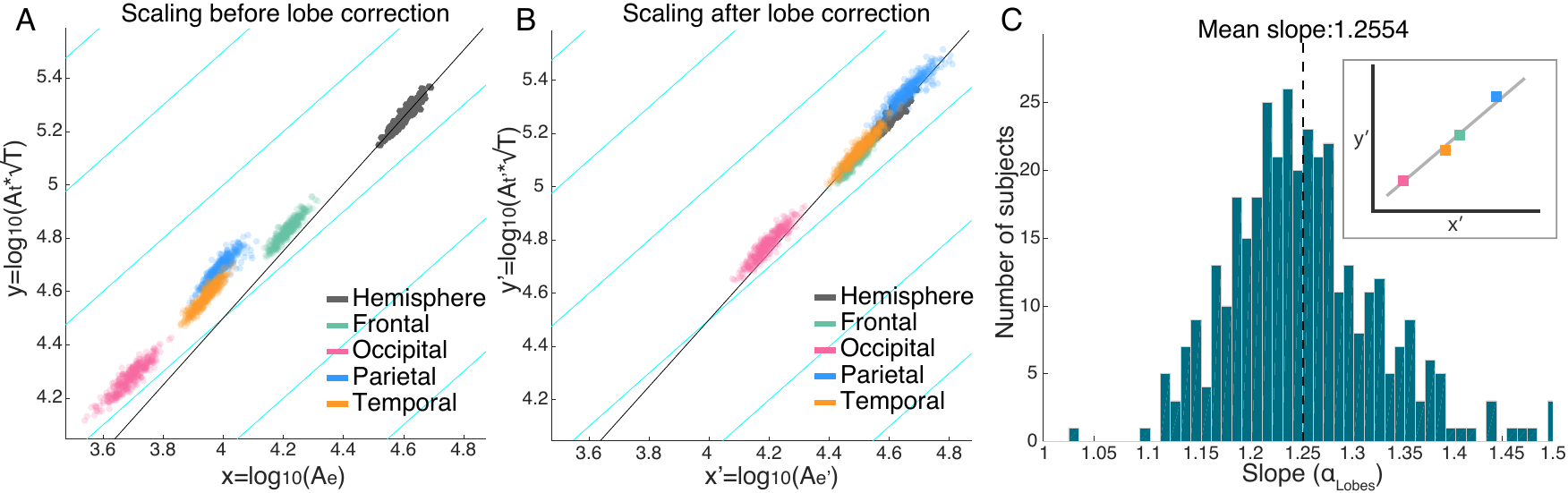}}
\caption{\textbf{Lobes in the same brain follow the same scaling law.}
\textbf{(A)} Scaling behaviour is shown for raw data of lobes in different colours, and for the whole hemisphere in grey. Grey line shows the linear regression for the whole hemisphere (slope=1.2557). Cyan lines indicate contour lines of constant gyrification index in this space (slope=1), along which our correction term operates. \textbf{(B)} Same as (A), only with correction term applied to the lobes to reconstruct their equivalent whole hemisphere data points. \textbf{(C)} Histogram of subject-specific regression slopes of the lobes. Dashed black line indicates the mean slope across subjects (being 1.2554). Inset exemplifies how the slope for a specific hemisphere was obtained through regression across its four lobes. For all panels, the HCP data was used, and we only included data for age 22-25 in this figure.
\label{Fig1_ScalingCorrection}}
\end{center}
\end{figure}

Fig.~\ref{Fig1_ScalingCorrection}A shows how the lobes naturally disaggregate according to their size and thickness in a $x=log(A_e)$ and $y=log(A_t\sqrt{T})$ plot. As expected, the lobes separately do not follow the universal relation (Eq.~\ref{UnivLaw}) for gyrification. All lobes are shifted to the bottom left in this plot compared to the whole hemisphere, and the shift is more pronounced for smaller lobes (e.g. occipital lobe) due to larger decreases in surface areas.

To arrive at measures that can be used to evaluate the scaling law of different lobes in the same cortex, we derived estimates of $A_e'$ and $A_t'$ by applying a correction term to the $A_e$ and $A_t$ of the partitions/lobes (see Methods Suppl. E for details). We obtained the estimates of $A_e'$ and $A_t'$ of each lobe, and left the average thickness $T$ of the lobe unaltered. These quantities can now be compared directly to each other by plotting $x'=log_{10}(A_e')$ against $y'=log_{10}(A_t' \sqrt{T})$, and to the original, whole cortex/hemisphere\footnote{We use the terms `cortex' and `hemisphere' interchangeably in this work.} (Fig.~\ref{Fig1_ScalingCorrection}B).

In Fig.~\ref{Fig1_ScalingCorrection}B, it is visually clear that the correction maps the data for all lobes (coloured data points) very close to the universal scaling behaviour of the whole hemisphere (grey data points, and grey line). Note that the correction moves the data points along the direction of constant gyrification index (cyan lines), as the correction term preserves the gyrification index of each lobe. In some cases, notably in the parietal/occipital lobe, the corrected areas can be larger/smaller than that of the whole hemisphere, which is a consequence of different lobes having gyrification indexes that are either lower or higher than the whole cortical average.

When we restrict our analysis only to lobes belonging to the same hemisphere, it becomes clear that, even when different lobes are dissimilar, taken together they also follow a single scaling rule (Fig.~\ref{Fig1_ScalingCorrection}(C) inset). In Fig.~\ref{Fig1_ScalingCorrection}(C) a distribution of slopes is obtained for the sets of four lobes in each hemisphere separately. Note that this derives a scaling exponent for each hemisphere separately (denoted as $\alpha_{Lobes}$), and is no longer a group-based estimate across different cortices (we denote the group based estimate of slope as $\alpha_{Hemispheres}$). Also note that there were no differences between males and females (p$>$0.05 in a ranksum test), hence we present all our data by grouping both sexes together. The average slope across the group is 1.2554, i.e. very close to the predicted value of 1.25. In Suppl. Text G we also performed this analysis using a linear mixed effect model, effectively estimating a joint slope across different cortices. The results show that the subjects do not have a significantly different slope to impact the group slope estimation (p=0.52), and that the 95\% confidence bounds around the group mean of the slope are +/- 0.01 (much smaller than the variance in the distribution shown in Fig.~\ref{Fig1_ScalingCorrection}C). \\

\subsection{Slopes across lobes does not change over age}

\begin{figure}[hbt]
\begin{center}
\centerline{\includegraphics[width=18cm]{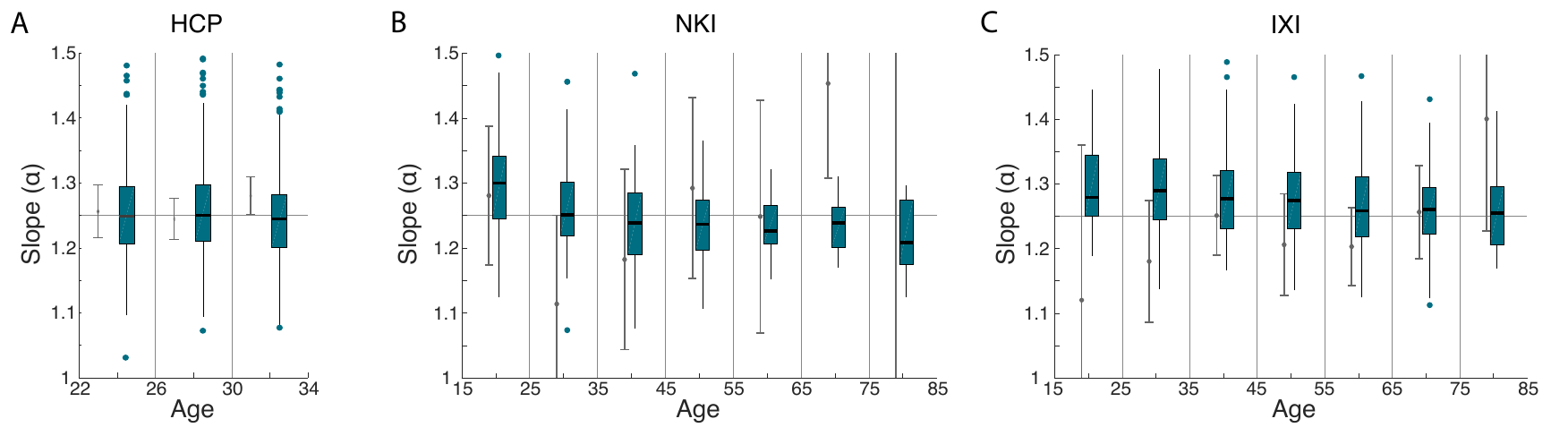}}

\caption{\textbf{Individual hemisphere slope estimates do not change over age.}
\textbf{(A)} Distribution of slope estimates ($\alpha_{Lobes}$) from lobe based regression is shown as box plots (dark green) over different age groups for the HCP data. For comparison, the group slope estimate ($\alpha_{Hemispheres}$) is shown as light grey errorbars (95\% CI). \textbf{(B,C)} Same as in (A), but for the NKI and IXI data, which span a wider age range. Suppl. Text B shows the raw data underlying this plot. 
\label{Fig2_SlopeAge}}
\end{center}
\end{figure}

To test if the individual slopes ($\alpha_{Lobes}$) change over age, we repeated the analysis from Fig.~\ref{Fig1_ScalingCorrection} for different age ranges (Fig.~\ref{Fig2_SlopeAge}). To enable comparison to the group slope estimate ($\alpha_{Hemispheres}$) we also used age categories. For the HCP dataset, there were enough subjects to allow for a four-year age categorization. For all other datasets ten-year categories were used. In all cases, the distribution of individual slopes $\alpha_{Lobes}$ cluster around 1.25 within one standard deviation. The raw data with continuous age is shown in Suppl. Text B. There is, however, the trend of a drift of median slope values with age, where a steeper slope is observed for younger subjects, and a shallower slope is seen for older subjects. We return to this observation in Discussion. It is also worth noting that $\alpha_{Lobes}$ estimates generally align better with 1.25 (median is closer to 1.25) than $\alpha_{Hemispheres}$, the slope estimates based on the full cortex (grey lines in Fig.~\ref{Fig2_SlopeAge}, which are the same slope estimates as in our previous publication \cite{Wang2016}). This is not surprising, given that the group slope estimates are directly influenced by outliers caused by e.g. segmentation and surface reconstruction errors \cite{Wang2016}. In lobe-based estimates, however, such outliers have little effect on the median of the distribution.

\subsection{Offset for all lobes decrease with age, as for whole cortex}

\begin{figure}[hbt]
\begin{center}
\centerline{\includegraphics[width=18cm]{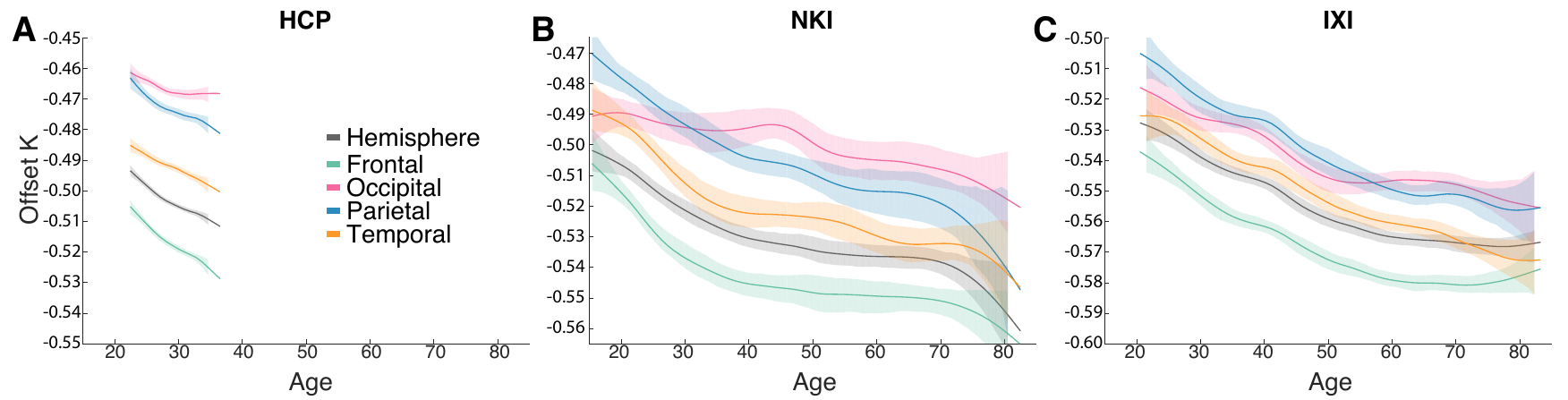}}
\caption{\textbf{Offset changes over age for different lobes and the whole hemisphere.}
\textbf{(A)} Offset ($K_{Lobe}$) is shown to change over age for different lobes for the HCP (900 subject release) data in four different colours. Note that shaded areas indicate 95\% bootstrapped confidence intervals for the mean, not the variance. For comparison, the offset for the whole hemisphere ($K_{Hemisphere}$) is shown in black. \textbf{(B,C)} Same as in (A), but for the NKI and IXI data, which span a wider age range. Suppl. Text B shows the raw data underlying this plot.
\label{Fig3_OffsetAge}}
\end{center}
\end{figure}

Based on the previous result, we assume a slope $\alpha_{Lobes}$ for each cortex of 1.25 across all ages in all datasets. This allows us to calculate the offset $K_{Lobe}=log(k')=log(A_t' \sqrt{T})-\frac{5}{4}log(A_e')$ separately for each lobe and each cortex. Fig.~\ref{Fig3_OffsetAge} shows the offset $K_{Lobes}$  over age for all four lobes, and also the entire hemisphere (where $K_{Hemisphere}=log(k)=log(A_t \sqrt{T})-\frac{5}{4}log(A_e)$) for reference. Overall, the offset decreases with age for all lobes, as expected from our results for the hemisphere \cite{Wang2016}. Generally, the lobes show the same rate of decrease as the hemisphere, although there is a hint of the occipital lobe decreasing at a slightly slower speed. Indeed when testing the slope of the decrease (k\~age) for differences between the lobes, significant differences were found (Suppl. Text A).

We additionally note systematic differences in offset between the lobes and between different datasets. We return to this observation in Discussion.

\subsection{Scaling of lobes in Alzheimer's disease is not altered}

\begin{figure}[hbt]
\begin{center}
\centerline{\includegraphics[width=18cm]{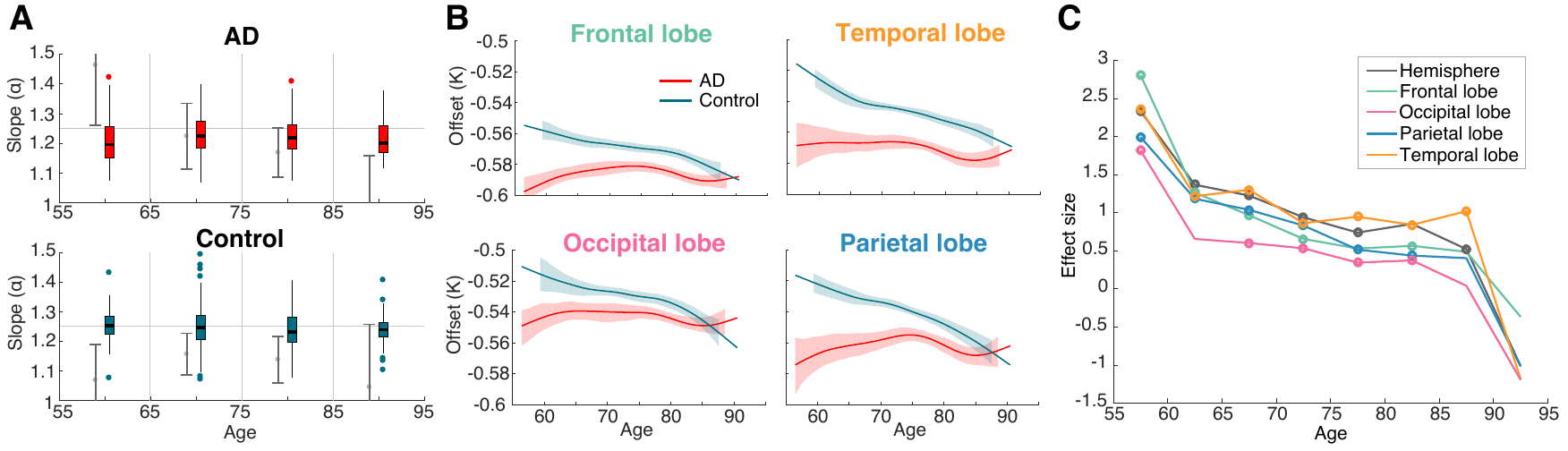}}
\caption{\textbf{Slope and offset over age for Alzheimer's subjects.}
\textbf{(A)} Slope estimates based on different lobes of the same hemisphere (box plots) for Alzheimer's patients (AD) and controls. As a reference the horizontal grey line indicates the predicted 1.25 slope. To enable comparison, we also show the group-based slope estimates as grey errorbars.
\textbf{(B)} Offset is shown for AD and control groups for different lobes. Solid line indicates the mean, and shaded area shows the 95\% bootstrapped confidence intervals for the mean.
\textbf{(C)} Effect size of offset ($K_{Lobe}$) from the control vs. AD between-group comparisons at different lobes. Empty circles show where p$<$0.05 in a ranksum test.
\label{Fig4_ADNI}}
\end{center}
\end{figure}

To enable further comparison of our results here with our previous results on the whole cortex scaling, we also investigate the same Alzheimer's dataset we analysed in \cite{Wang2016}. The dataset is a cross-sectional cohort of about 200 Alzheimer's disease (AD) patients and 200 healthy controls in the same age range. In terms of the slope of the scaling law, we previously found that a significant gender specific change may occur in Alzheimer's disease (AD) compared to controls. We re-investigated this by deriving a distribution of slopes from how the lobes scale within the same cortex. Fig.~\ref{Fig4_ADNI}~(A) shows that the previously found difference in $\alpha_{Hemispheres}$ disappears with this estimation of slope based on the lobes ($\alpha_{Lobes}$). 

We previously also noted that even the slopes in the control cohort seemed to be decreasing, which also is not the case with our lobe based analysis. All control slope distributions are centred on 1.25, without any significant changes with age (Fig.~\ref{Fig4_ADNI}~A).

In terms of offset (calculated again assuming a slope of 1.25), we observe the same trend for the lobes as with the whole hemisphere: $K_{Lobe}$ for the AD group remains low and stationary for all ages, while the control group slowly drifts down to this level over age (Fig.~\ref{Fig4_ADNI}~B). There are, however, some subtle but significant differences between the lobes. When measuring the difference between AD and control group (as effect size between the two distributions), the occipital/temporal lobe is shown to have the smallest/biggest effect size overall (Fig.~\ref{Fig4_ADNI}~C). The effect sizes also show a sharp decrease overall after age 60, then stays roughly constant until age 85.

\section{Discussion}

In summary, we demonstrated that lobes of the same cortex follow the same universal scaling law as different human cortices, and as different cortices of different mammalian species. The universal scaling of lobes holds for healthy human brains of different ages and sexes, and even for subjects with Alzheimer's disease. The offset parameter, which we associated previously with axonal mechanical tension, also decreases with age in each lobe in a similar rate as for the whole cortex. In Alzheimer's disease, cross-sectionally, we observe that the difference in offset between patients and controls is largest in the temporal lobe, smallest in the occipital lobe, and stays roughly constant between age 60 and 90 for all lobes.

\textbf{Methodological challenge:} In order to arrive at a slope estimate for each individual cortex ($\alpha_{Lobes}$), we partitioned the cortex into four different lobes, and applied a correction term to the total and exposed surface area of the lobes. The correction term is derived from the proportion of Gaussian curvature located on each lobe, as a topologically invariant measure of size \cite{Hazewinkel2013encyclopaedia}. However, we acknowledge that the estimation of Gaussian curvature from a surface is technically non-trivial, and numerically challenging (see Suppl E for details). We believe that the numerical errors (which generally increase with the size of the lobe) have led to a systematic difference in the offset of each lobe, giving bigger lobes a smaller offset, and smaller lobes a bigger offset (Fig.~\ref{Fig3_OffsetAge}). The numerical effects may also have influenced the finding that our slope estimates seem to decrease very slightly with age (Fig.~\ref{Fig2_SlopeAge}), as the size of the brain, and hence also the size of the lobes, is slightly smaller in older age on average \cite{Wang2016}.

\textbf{Slope estimates:} Apart from the potential numerical inaccuracies, the slight slope ($\alpha_{Lobes}$) decrease may also be driven by a genuine biological effect. It is conceivable that with a slope estimate for individual cortices, we are more sensitive to small changes with age on a more local level of the brain. In support of this, Madan \textit{et al.} recently reported in a cross-sectional study a decrease in fractal dimension of the cortical ribbon with age \cite{Madan2016}. If the scaling law is indeed true for parts of the brain, as for the whole brain, then the fractal dimension of the cortical ribbon should be 2.5, which is very close to the reported values. Both a decrease of fractal dimension with age, and similarly our decrease of slope ($\alpha_{Lobes}$) with age, may then suggest a genuine biological effect indicating that the scaling behaviour may change over age.

However, a closer observation of our data shows that the decrease in slope is not consistent in absolute values across data sets. In Suppl. C we show that both our slope, as well as the fractal dimension are strongly correlated with our offset $K_{Hemisphere}$ on an individual hemisphere basis. It is difficult to draw definite a conclusion at this stage, but we speculate that rather than a genuine slope change, the observed effect here (and maybe even in \cite{Madan2016}) are due an secondary effect/scale that is currently assimilated in our offset parameter $K_{Hemisphere}$. Future work may be able to resolve this by finding a reliable way of estimating the scaling behaviour within a cortex without being confounded by factors such as size of the cortex (or partition). Nevertheless, we can conclude from our analysis so far that lobes of the same cortex also obey the same universal scaling law as different cortices from the same species, and different mammalian species.

\textbf{Offset $K$ estimates:} In terms of our scaling law, we previously interpreted a drop in $K$ over age as a decrease in tension along white matter axons\cite{Sack2009}. However, there are other effects that would also change the value of $K$ over time according to our proposed mechanism. Notably, a decrease in axonal cross-sectional density in the white matter (possibly indicated by diffusion imaging studies \cite{Cox2016}) would have the same effect as a proportionally equal decrease in axonal tension.  Intracranial pressure changes would also affect the value of $K$. For the lobes, $K_{Lobe}$ generally follows the overall behaviour of the whole cortex. Relative to each other, the frontal and temporal lobes experience the largest rate of decrease overall, and the occipital lobe appears least affected by age (Fig.~\ref{Fig3_OffsetAge}). These results mirror findings in the literature of cortical thickness changes over age \cite{Mcginnis2011,Hogstrom2012,Fjell2015,Walhovd2016}. The differential effect of age on cortical thinning has also been coined the `last in, first out' effect. The primary sensory and motor areas develop first prenatally, and the association areas follow with a slight delay and continue growing post-natally. The association areas are earliest affected by cortical thinning in aging, and the primary sensory and motor areas are only affected later on \cite{Mcginnis2011,Oh2014}. In the light of our results on the lobes, the differential effect of age on $K$ could then be similarly interpreted as a `last in, first out' effect in tension (e.g. caused by changes in axonal density) over age.


\textbf{Alzheimer's disease (AD):} We found no significant slope ($\alpha_{Lobes}$) changes with age, or gender in either AD, or control groups, for slope estimates derived from the lobes; and no overall difference to 1.25. This finding contrasts with the results we obtained previously based on regression across a population \cite{Wang2016}, where we detected a slope change with age in the AD group (Fig.~\ref{Fig4_ADNI}A, grey error bars). We suggest that the lobe based estimations may be more reliable and overcome some of the sample size and outlier issues in our previous work. Additionally, systematic regional variations on the cortex may have also impacted our previous results, due to the heterogeneity of cortical atrophy patterns in AD. With the lobe-based slope estimates, there might be a slight trend of a systematically shallower slope in AD than controls, although again future work will have to show if this is a genuine characteristic of the disease or a side effect of our method. At this stage, however, we conclude that there are no slope changes in AD or controls, and all slope estimates agree with the theoretically predicted 1.25.

The changes in offset $K_{Lobe}$ over age that we found in the lobes (Fig.~\ref{Fig4_ADNI}B,C) agree with the overall picture we obtained previously for the whole cortex \cite{Wang2016}. We also observed in our cross-sectional study no differential progression between the lobes in $K_{Lobe}$ (and thickness, see Suppl. D) with age in AD. Morphological changes in AD are primarily observed as reductions of grey matter volume and cortical thickness \cite{Pini2016}, which are believed to be triggered by targeted neuronal disconnections \cite{Delbeuck2003,Bartzokis2004}, leading to an accelerated brain aging process \cite{Lorenzi2015}. Indeed, the low and almost constant $K$ in AD patients over the entire age range, agrees with these observations of accelerated aging caused by the AD pathology. In this regard, the occipital lobe shows least difference between AD and controls overall (Fig. 4B,C). This agrees with findings in the literature showing that AD seems to least affect the occipital lobe \cite{Dickerson2009a,Lemaitre2012,Blanc2015,Pini2016}, compared with other lobes and its sub-structures such as the temporal (e.g. middle temporal gyrus) and parietal (e.g. precuneus cortex) lobes \cite{Dickerson2016,Frisoni2007}. We have stated previously that the consistently low value of $K$ in AD, in contrast to its gradual decrease in healthy aging, suggests that a premature morphological aging of the cortex may be seen as an aspect of AD. On that note, it is interesting that the occipital lobe, the one least affected by AD, is also the one least affected by aging in terms of $K$.

As for the cause of the low value of $K$ in AD, it is hard to say anything definitive without a longitudinal analysis to track how $K$ change for individuals (and, in particular, if there is a sudden decrease of $K$ at or just before the onset of AD, an effect that could have important prognostic value). However, a decrease in axonal density, something we expect would lead to a decrease in $K$, has recently been associated with AD in the literature \cite{Slattery2017}.


\textbf{Conceptual advances:} We showed in our previous work \cite{Wang2016} that the proposed scaling law may be able to offer a new set of three natural morphological variables, that are directly and independently related to (1) cortical thickness, (2) brain size, and (3) mechanism of folding. Importantly, in this coordinate system we can disentangle the three effects from of each other, so that e.g. brain size does not influence the measure of folding (unlike the gyrification index, which is influenced by  brain size). However, our previous work was limited by the fact that the scaling law could only be applied to a cohort, meaning that the folding mechanism within an individual hemisphere could not be studied.

In this work, we have proposed a way to extend these new variables to partitions of the cortex, enabling an analysis of folding of an individual brain. This is significant step forward, as it now allows us to analyse mechanisms of folding within subjects, rather than just for a group of subjects. With this extension, we were able to show that parts of the same cortex still follow the proposed universal scaling law, and hence still can be projected and understood in our natural coordinate system. Conceptually, this also confirms our hypothesis that the mechanism of folding is universal across mammalian species, within the human species, and even within a single cortex. Future work will try and resolve the small residual dependencies on partition size we still observe, and fully develop our method into a local measure to enable its use in e.g. clinical applications for both within and between subject studies.

\section{Methods}

\subsection{MRI and data processing}
We included a full overview of the MRI data we used in Suppl. Text F. Briefly, the datasets are obtained from four sources: The HCP data \cite{VanEssen2012} we obtained is from the 900 subjects release. For the IXI data \cite{IXI} we only used the Guy's Hospital subdataset (site with biggest sample size). For the NKI Rockland Sample and ADNI\footnote{Data used in preparation of this article were obtained from the Alzheimer’s Disease
Neuroimaging Initiative (ADNI) database (adni.loni.usc.edu). As such, the investigators
within the ADNI contributed to the design and implementation of ADNI and/or provided data
but did not participate in analysis or writing of this report. A complete listing of ADNI
investigators can be found at:
\url{http://adni.loni.usc.edu/wp-content/uploads/how_to_apply/ADNI_Acknowledgement_List.pdf}} data, we used exactly the same subdataset as in our previous publication \cite{Wang2016}.

\subsection{Partitioning of cortex}

There are numerous ways to partition a cortex, and in principle the method outlined below should be applicable to any choice of segmentation as long as the partitions are not too small (i.e. including several gyri and sulci). A natural partition, however, is dividing each hemisphere in four lobes (parietal, occipital, frontal and temporal), the definition of each being consistent across individual humans \cite{Fischl2002}.

To partition the cortex into lobes, we used the Freesurfer Desikan-Killiany parcellations and followed the Freesurfer lobe assignments (including the assignment for the cingulate) \cite{FSwebsiteparcel}. See Suppl. E for more details.

\subsection{Correction term}

For all closed surfaces (or, approximately, for almost closed surfaces, such as a cortical hemisphere intersected by the corpus callosum), the integral of the Gaussian (or intrinsic) curvature $G$ is always $I_{G}=\oint G dA=4 \pi$ \cite{Hazewinkel2013encyclopaedia}. This so-called topological invariant is thus a natural measure of the relative sizes of different partitions of a closed surface, that is insensitive to deformations and details about shape. It is also insensitive to transforming a smooth surface (such as an actual cortical surface) into a triangulated one (such as the computational representation thereof). In the latter case, the Gaussian curvature becomes concentrated on the vertices, and in each vertex it is simply the difference between $2 \pi $ and the sum of the angles impinging on it\cite{Mesmoudi2012}, which is easily computed. Now consider a partitioning of the cortex into four lobes in both the total and exposed surfaces. The total Gaussian curvature of teh whole cortex ($4\pi$) will be distributed among the partitions. 

While reconstructing a full cortex from a partition, we require that certain properties remain unchanged: The average thickness of the reconstructed cortex $T'^P$ should be the same as that of the partition $T^P$. The gyrification index $g^P=\frac{A_t^P}{A_e^P}$ should also not change. Finally, the average Gaussian curvatures for the total and exposed areas in the partition, $\bar{G}_e=\frac{I_{G}^P}{A_e^P}$ and $\bar{G}_t=\frac{I_{G}^P}{A_t^P}$ should be preserved, with $I_{G}^P=\int_{A_e^P} G dA$ being the integrated Gaussian curvature of the partition, and $A_e^P$ the surface area of the partition\footnote{Mathematically, the Gauss-Bonnet theorem ensures that we would obtain exactly the same value using $A_t$ instead of $A_e$; in practice, using $A_e$ makes our results less sensitive to the precise placement of partition boundaries}. Hence:

\begin{subequations}\label{AreaIndependentVars}
    \begin{align}
        T'^P &= T^P,\\
        g'^P &= \frac{A_t'^P}{A_e'^P}=\frac{A_t^P}{A_e^P}=g^P,\\
        \bar{G}_e'^P &=\frac{4\pi}{A_e'^P} = \frac{I_{G}^P}{A_e^P} =\bar{G}_{e}^P,\\
        \bar{G}_t'^P &=\frac{4\pi}{A_t'^P} = \frac{I_{G}^P}{A_t^P} =\bar{G}_{t}^P.
    \end{align}
\end{subequations}
From which follows the corrected total and exposed areas for the partition, now with size effects removed and directly comparable to each other and to the whole cortical hemisphere,

\begin{equation}\label{AreaIndependetVars21}
   A'^P_e = \frac{4 \pi}{I_{G}^P} A^P_e   \;\;  \textrm{and}  \;\;      A'^P_t = \frac{4 \pi}{I_{G}^P} A^P_t.
\end{equation}

We describe the reasoning behind the correction term and estimation of $I_{G}^P$ in more detail in Suppl. Text E.

\subsection{Correction in slope estimate}
To remove the systematic Gaussian error due to the dispersion in the value of $I_G^P$, let $\sigma^2_{\log{I_{G}^P}}=\sigma^2_{I_{G}^P}/(\bar{I}_{G}^P)^2$ be the variance in the logarithm of the correction term $\log{I_G^P}$, and $\sigma^2(\log A_e'^P,\log A_t'^P\sqrt{T^P})$ the covariance matrix for different partitions of the same hemisphere. Then it can be shown (Suppl. Text E) that the slope across lobes $\alpha_{Lobes}$ is given by $\alpha_{Lobes} = \frac{1}{\gamma+\sqrt{1+\gamma^2}}$, where $\gamma = \frac{\sigma^2_{11}-\sigma^2_{22}}{2\sigma^2_{12}-\sigma^2_{\log{I_{G}^P}}}$.

\section{Acknowledgements}
Data were provided in part by the Human Connectome Project, WU-Minn Consortium (Principal Investigators: David Van Essen and Kamil Ugurbil; 1U54MH091657) funded by the 16 NIH Institutes and Centers that support the NIH Blueprint for Neuroscience Research; and by the McDonnell Center for Systems Neuroscience at Washington University. L.P.R. is supported by National Institute of Health Research (NIHR) Biomedical Research Centre (BRC) at Newcastle University.

We thank Kathryn Garside for discussions, and Andre Muricy for code that inspired the current analysis for partitioning exposed surface areas.

\bibliographystyle{ieeetr}
\bibliography{main}


\end{document}